\documentstyle[twocolumn,aps]{revtex}

\newcommand{\beq}{\begin{equation}}
\newcommand{\eeq}{\end{equation}}
\newcommand{\beqa}{\begin{eqnarray}}
\newcommand{\eeqa}{\end{eqnarray}}
\newcommand{\beqar}{\begin{eqnarray*}}
\newcommand{\eeqar}{\end{eqnarray*}}

\def \la {\langle}
\def \ra {\rangle}

\begin{document}

\title{ \bf
\Large Local/Non-Local Complementarity in Topological Effects}

\author{   Yakir Aharonov$^{(a,b)}$,
           Benni Reznik$^{(a,\dag)}$
\\
$^{(a)}$ {\em  \small School of Physics and Astronomy, Tel Aviv
                      University, Tel Aviv 69978, Israel.}\\
$^{(b)}$ {\em \small     Department of Physics,
          University of South Carolina, Columbia, SC 29208.}
}

\date{\today}
\maketitle

\begin{abstract}

{
In certain topological effects 
the accumulation of a quantum phase shift is 
accompanied by a local observable effect.
We show that such effects 
manifest a complementarity between non-local and local
attributes of the topology, which is reminiscent but yet different
from the usual wave-particle complementarity.
This complementarity is not a consequence of non-commutativity,
rather it is due to the non-canonical nature of the observables. 
We suggest that a local/non-local complementarity  
is a general feature of topological effects that are  ``dual''
to the AB effect.
}

\end{abstract}
---------------------------------------------------------------

In the Aharonov-Bohm (AB) effect \cite{AB,PT}
a charge moves around a magnetic flux filament
in a region with vanishing electromagnetic fields.
The charge experiences no electromagnetic forces, 
yet it accumulates a topological quantum phase shift.
Topological effects which are ``dual'' to the AB effect
have been discovered for neutral particles.
Aharonov and Casher \cite{AC,ACexp,Allman} 
have shown that particles carrying a magnetic 
moment and moving around a 
 straight  wire with uniform charge density,
will experience no force
but acquire a phase shift analogous to the AB phase 
\cite{APV,Goldhaber89,Anandan89,Hagen}.
More recently, it has been shown by 
He and McKellar \cite{HM} and by Wilkens \cite{Wilkens},
that a neutral particle carrying an electric dipole 
also exhibit similar  dual topological effects  \cite{Franson,more}.
Nevertheless, unlike the AB effect,  
in these dual cases the local fields along the trajectory of the particle
do not vanish. Consequently, as was pointed out by
Peshkin and Lipkin \cite{Peshkin,PL}, 
the accumulation of the attendant quantum phase shift may be 
accompanied  by a local observable effect.

In this letter we suggest that such ``dual'' topological effects 
manifest a complementarity between the non-local and the local
attributes of the topology, which is reminiscent but yet different
from the usual wave-particle complementarity;
by measuring a local observable we disturb the
non-local phase information on the topology.
However the complementarity suggested here is not a consequence 
of non-commutativity,
rather it is due to the non-canonical nature of the corresponding
observables \cite{AS}.

To illustrate this complementarity let us begin with the 
Aharonov-Casher (AC) effect \cite{AC}. 
In the AC effect the magnetic moment $\mu$ interacts with 
the electric field $\vec E$, the vector
potential-like
term,  $\vec\mu \times \vec E$. 
which induces phase a
\beq
\phi_{AC} = {1\over \hbar} \oint {\vec\mu\times\vec E }\cdot \vec dl=
{\mu\lambda\over \hbar} n_{AC} .
\eeq
Here $\mu$ is the projection of $\vec\mu$
in the direction of the line,
$\lambda$ is the charge per unit length, and  $n_{AC}$ is the winding number
of the  magnetic moment's trajectory around the line.

We note that several similarities exist between the AC and AB effects.
Since in both cases the force vanishes, they are ``force-free'' effects 
\cite{APV}.
Moreover, in both cases a ``vector-potential'' coupling
gives rise to a topological phase;
indeed for a closed trajectory, the  AC phase and the AB phase, 
are insensitive to the details of the path, 
and are determined from the winding number alone.

However this similarity breaks down in one important aspect \cite{Goldhaber89}.
Since in the AB effect the particle couples to a gauge
field, the locally accumulated phase is not gauge invariant.
Only the phase accumulated in an interference experiment, 
on a closed loop, is a gauge invariant quantity.
Therefore, the AB effect is sometimes described as being non-local.
On the other hand, since in the present case the magnetic moment clearly
couples directly to the field strengths $\vec E$,
the  locally accumulated phase is a gauge invariant 
and meaningful observable.

This naturally raises the question of the non-locality of 
such effects\cite{RA,Goldhaber-charge,interplay}. 
In particular, Peshkin and Lipkin \cite{PL}  
have noted that a magnetic field is present in 
the local reference frame of the magnetic moment. Consequently,
the magnetic moment vector precesses around the direction of the 
magnetic field by an angle that turns out to be proportional to the phase
shift.
Since the accumulation of the quantum phase shift is accompanied by a local 
precession, and since the latter is locally observable, they
concluded that the AC  
is inherently local.

It was however implicitly assumed that the 
accumulated phase and the precession are  two  
attributes of the effect, which are simultaneously meaningful.
Indeed the autocorrelation operator suggested in \cite{PL}
commutes with the phase operator (that we define below). 
This apparently implies that we can locally measure the
rotation of the magnetic moment without disturbing the accumulated phase.
Nevertheless, the precession and the accumulated phase are 
non-canonical variables \cite{AS}, and  
{\em for non-canonical variables  commutativity 
does not imply  mutual observability}. 
In fact, we will show that by observing the local 
precession we necessarily induce an uncertainty in the accumulated phase, 
in a similar way  as for ordinary canonical conjugate variables.

Let us first show that the above complementarity
is required for consistency of the AB effect with
ordinary wave-particle complementarity \cite{FR}. 
Consider the usual AB interference experiment of charged particles
around a solenoid enclosing a magnetic flux.
We wish however to regard the effect of the charge on the 
internal degrees  of freedom of the fluxon.
For simplicity let space be only 2-dimensional, hence
the solenoid is replaced by a magnetic moment $\vec \mu$
pointing in the ``up" direction. 
The magnetic moment generates an AB vector
potential, hence the charge  acquires the usual AB phase shift
that can be observed in an interference experiment.
On the other hand, consider now the effect of the charged particle on the 
magnetic moment. In the rest frame of the magnetic moment, the 
moving charge generates the magnetic field 
$\vec B = \vec v\times \vec E$, where $\vec v$ is the velocity of the
charge and $\vec E$ its electric field. 
As noted above, this magnetic
field causes a precession of the magnetic moment $\vec\mu$.
By evaluating the angle of precession $\delta\varphi$
when the charge $q$ is moving along either the upper or lower
side of the magnetic moment we find $\delta\varphi \propto \phi_{AC}$
(see below).
Therefore, by measuring $\delta\varphi$ we can determine 
on which side of the fluxon the particle moved.
Clearly, that contradicts the wave-particle  complementarity principle, 
unless the measurement of precession 
 destroyed the coherence of the two  trajectories.

We shall next examine this process in more detail
to see how in actuality this loss of coherence happens. 
The non-relativistic Hamiltonian in 3-dimensions \cite{AC,Anandan89,Hagen}
\beq
H= {(\vec P + \vec \mu\times \vec E)^2\over 2m} - {\mu^2E^2\over 2m}
\label{H}
\eeq
describes a spin-half 
neutral particle carrying  magnetic moment
$\vec\mu = \mu \vec \sigma$,
which moves in an
electric field $\vec E$.
In the AC effect, the electric field
is generated by a straight  wire with uniform charge density $\lambda$. 
If the particle moves in the plane orthogonal to the wire, and 
the momentum in the direction of the wire vanishes,  
one finds \cite{Hagen} that $H$ 
effectively reduces to a 2-dimensional Hamiltonian.
Using polar coordinates  $(r(x,y),\theta(x,y))$ 
where the  charged wire is located at $r=0$, and points in the $z$
direction,
we get 
\beq
H= {p_r^2\over2m} + { (p_\theta + \xi \sigma_z)^2\over2m
  r^2}  ,
\eeq
where $\xi \equiv \lambda\mu/2\pi$.

The Heisenberg equations of motion for the spin 
\beqa
\dot\sigma_x &=& 
 - {2\xi\over \hbar mr^2} p_\theta\sigma_y \nonumber\\ 
 \dot\sigma_y &=&  {2\xi\over \hbar mr^2}p_\theta \sigma_x \nonumber \\ 
\dot\sigma_z&=&0 .
\label{eq}
\eeqa
describe a precession of the spin around the $z$ axis. 
When the magnetic moment moves 
between time $t_0$ to time $t$
along a path joining points with angular coordinates $\theta(t_0)$
and $\theta(t)$, we find that 
(up to the trivial phase ${2\xi\over\hbar m} \int^t_{t_0}{dt'\over r^2}$),
the precession is generated by the unitary operator
\beq
U(t,t_0)= e^{-i{2\xi\over \hbar }\int_{t_0}^t \sigma_z{\dot\theta(t')}dt'} 
\eeq
Indeed, this precession is induced by 
the magnetic field, $B_z=(\vec v\times\vec E)_z=2\xi\dot\theta$, 
experienced by the spin in its rest frame.
If $\sigma_z$ is constant, the angle of precession $\varphi$ is 
hence: $\varphi={2\xi }\sigma_z(\theta(t)-\theta(t_0))/\hbar$. 

Consider now the wave function $\psi$ of the magnetic moment. 
For the above trajectory, it changes according to 
\beq
\psi \to  U(t,t_0)\psi = e^{-i\delta\phi_{AC}}\psi
\label{psi}
\eeq
where
\beq
\delta\phi_{AC}(t,t_0) = {\xi\over\hbar} \int^t_{t_0} 
\sigma_z  {\dot\theta(dt')} dt' .
\label{phase}
\eeq
In the AC effect, $\psi$ is an eigenstate of $\sigma_z$. 
If $\sigma_z=1$,  $\delta\phi_{AC}(t,t_0)=
\xi(\theta(t)-\theta(t_0)/\hbar$.
We will henceforth refer to $\delta\phi_{AC}(t,t_0)$  in (\ref{phase})
as the ``phase operator'' . 
In general it describes the phase accumulated along a definite
trajectory for  arbitrary  $\sigma_z$.

Let us next examine what is the effect on a system 
when the spin precession is measured.
The rotation implies that the spin at $t_0$ is related
to the spin at some latter time $t$. 
Particularly we have the  identity
\beq
C_{\varphi}(t,t_0)\equiv U^\dagger(t,t_0)\vec\sigma(t_0)U(t,t_0)
 - \vec\sigma(t) =0 ,
\label{identity}
\eeq
which follows from the equation of motion of the spin.
By observing  that $C_{\varphi}(t,t_0)$ indeed vanishes we can verify that
the spin rotates.
One  might think  that since
\beq
[C_{\varphi}(t,t_0) , \delta\phi_{AC}(t,t_0)]=0 ,
\label{commute}
\eeq
we should actually be able to observe simultaneously both
the  precession operator $C_{\varphi}(t,t_0)$ and the phase operator 
$\delta\phi_{AC}(t,t_0)$.

However the above commutativity is a dynamical result, i.e. 
it is valid only by virtue of equations of motion (\ref{eq}).
To define  $C_\varphi(t,t_0)$ and  $\phi_{AC}(t,t_0)$ 
one has to specify the Hamiltonian (\ref{H}).
For such non-canonical variables \cite{AS},  
non-commutativity 
does not imply mutual observability.

To observe $C_\varphi(t,t_0)$, we have to couple to the system
twice, first at time $t_0$ and then at a later time $t$.
Since the coupling of the system to a measuring device at time $t_0$, 
changes the Hamiltonian (\ref{H}) 
(because we must add to the Hamiltonian new terms
describing the interaction of the system with a measuring device),
the spin component $\sigma_z$ will no longer be constant, and
the accumulated phase (\ref{phase}) will change by an uncertain
amount. 

Let us examine in more detail the uncertainty produced in 
$\delta\phi_{AC}(t,t_0)$ when $C_\varphi(t,t_0)$ is measured.
To this end, we  couple at $t=t_0$ to 
$\sigma_i$ and at some later time $t$
to the rotated spin ${U^\dagger \sigma_i U}$.
To be able to observe a precession of a single spin, 
we must be sure that the spin has rotated by a 
sufficiently large angle, say $\varphi=\pi/2$.
Choosing $i=x$, we get for this case  
\beq
C_{\pi/2}(t,t_0)=\sigma_y(t)- \sigma_x(t_0) = 0 .
\eeq

Because $C_{\pi/2}(t,t_0)$ is of order unity it must be
observed with precision
\beq
\Delta C_{\pi/2}(t,t_0) \ll 1 .
\eeq
Hence  $\sigma_x(t_0)$ is effectively  measured with precision
$\Delta \sigma_x\ll 1$. During the time 
interval $(t,t_0)$, $\sigma_z$ then becomes uncertain by
$\Delta \sigma_z \approx 1$. 
The  consequent uncertainty in the AC phase 
\beq
\Delta\phi_{AC}(t,t_0) = {\xi\over\hbar}(\theta(t)-\theta(t_0))
 \Delta\sigma_z \approx {\pi/4}
\eeq
is hence sufficiently large to erase the phase information.
This achieves our goal of showing that by measuring the precession 
we destroy the coherence.

More generally, we will be able to infer that $C_\varphi(t,t_0)=0$
only statistically.
Consider for example the limiting case, that
the spin has precessed by only  a small angle
$\varphi\ll1$. Let $|x\ra$
be the eigenstate of $\sigma_x$, and denote the rotated
eigenstate by $|\varphi\ra$. 
Hence  $|\la\varphi|x\ra| \simeq 1 -\varphi^2$.
To verify  the precession,
one has to repeat the experiment over a sample of 
 $N\sim {1\over \varphi^2}$ spins, 
all initially in the same $|x\ra$ state,
and measure separately for each spin
the operator $C^i_\varphi(t,t_0)$.
The total phase accumulated by the $N$ spins, $\phi^N_{AC}
=\sum_{i=1}^N \phi_{AC}^i$, will become uncertain by
\beq
\Delta\phi_{AC}^N = {\xi\over\hbar} (\theta(t)-\theta(t_0))
 \sum_{i=1}^N \Delta\sigma_z^i
\approx \varphi \sqrt N/2 \sim 1/2 ,
\eeq
where the relation,  $\phi = 2\xi(\theta(t)-\theta(t_0))/\hbar$, was used, 
and we have assumed that the uncertainties $\Delta\sigma_z^i\approx 1$, 
for each spin, are independent.
Therefore, if we verify that the $N$ spins precess, the total 
accumulated phase  becomes uncertain. This verifies our claim 
also for this case.

Next consider a  special case 
where the spinor nature has a special role.
Suppose that  
the spin  rotates  around the $\hat z$ axis, by  either $\varphi=+\pi$
or $\varphi=-\pi$.
In both cases, of either a clockwise or a counter clockwise rotation,  
$\sigma_x$ changes to $-\sigma_x$ and  
\beq
C_\pi = \sigma_x(t) + \sigma_x(t_0)=0 . 
\eeq
In space-time these two alternatives  correspond to 
a magnetic moment moving along either a clockwise or a counter-clockwise 
path around the charged wire with  
$(\theta(t)-\theta(t_0))=\pm \hbar\pi/2\xi$.
 Since both paths give rise to the same 
rotation of the spin, they cannot be 
distinguished by measuring $C_\pi(t,t_0)$.
Therefore, in this particular case, consistency with ordinary 
wave-particle complementarity does not require that coherence must be lost. 
So it may appear that this provides a counter example to our 
claim.

However by observing $C_\pi$, we are still unable to distinguish between
a non-trivial or a  trivial phase 
in an interference experiment, i.e. we cannot detect a non-trivial
topological effect.
To see that, let us compare two cases:
first consider an AC effect where the charged line generates the phases,
$\phi_{AC} = \pi/2$ on one path and   $-\pi/2$  for the other.
This yield a relative non-trivial $\pi$ phase.
In the second setup we arrange a special 
charge distribution which generates 
the {\em same}  $\pi/2$ phase for {\em both} paths.
The relative phase in the second case is trivial, however, 
since the spin rotates by $\pi$, $C_\pi=0$ in both cases, 
and we cannot distinguish by performing this measurement 
between the cases of a  topological and 
a non-topological effect.

Similar reasoning applies for the case of 
a magnetic moment moving through a region with 
a homogeneous but time dependent magnetic field $B(t)$. 
The corresponding phase shift
\beq
\phi_{AC} = {1\over \hbar} \int \vec\mu \cdot \vec B(t') dt' 
\eeq
was observed by
Allman et. al. \cite{Allman}.
This effect 
is sometimes referred to in the literature as the Scalar-AB effect,
because the interaction term $\vec\mu \cdot\vec B$ in 
(\ref{H}) is  analogous to the $qV$ term in the AB Hamiltonian
$(p-qA)/2m + qV$, which gives rise to the potential-AB
effect. 
Since with the inclusion of a magnetic field the AC Hamiltonian is 
$H_{AC} = (p+\mu\times E)^2/2m - \mu\cdot B$,
perhaps it is more natural to identify this phase as 
a   ``potential-AC effect''.
It can be readily shown that  our arguments for the AC effect follow,
by replacing the particle's rest-frame magnetic field, $\dot\theta(t)$, 
with $B(t)$.

So far we have shown a complementarity relation in the cases of the AC 
and the potential AC (or Scalar-AB) effects. Nevertheless,  
the gedanken-experiment suggested earlier indicates  
that a similar complementarity  relation exists for 
other ``dual" effects.                  
Such is the topological effect for electric dipoles \cite{HM,Wilkens,more}, 
which is manifested via a ``vector-potential'' $\vec d \times \vec B$, 
or a ``potential" $\vec d\cdot \vec E$ \cite{Wilkens}.
It was noted that the effect for electric dipoles 
can be obtained from the AC setup by a  Maxwell duality transformation \cite{Franson}.
To connect our gedanken experiment with these cases, 
we will however make use of a different type of duality \cite{AC} 
\beq
{\rm charge} \longleftrightarrow {\rm magnetic\  moment}
\eeq
which transforms the magnetic filament and the charge in the AB effect
to a charged wire and a magnetic moment in the AC effect and vice versa.
This duality, is closely related to the Galilean invariance 
of the non-relativistic charge-magnetic moment system.
The total accumulated phase
depends only on the relative motion of the charge and the magnetic moment. 
Hence by a duality transformation we transform 
from the rest frame of the magnetic moment (in the AB effect) to the 
rest frame of the charge (in the AC effect).
As we already have shown, since the phase is ``common'' to the charge and the magnetic moment, 
the consistency of the AB effect with ordinary wave-particle 
complementarity necessitates a local/non-local complementarity for 
the ``companion" dual effect. 

We next note that the  ``potential" electric-dipole effect 
generated by  the $\vec d\cdot \vec E$ term \cite{Wilkens}, 
is dual to the non-local potential AB effect:
let two charged plates of a  ``capacitor"  initially overlap 
each other. A potential difference
between the two sides of the capacitor is then formed 
when a charged particle passes close to the capacitor 
(so that $\vec E\simeq 0$), 
by changing temporarily the distance between the plates.
The charge then experiences no force but accumulates the 
AB phase ${q\over\hbar}\int V(t')dt'$. 

Consider now the duality transformation 
\beq
{\rm charge} \longleftrightarrow {\rm electric\  dipole}
\eeq
which replaces the capacitor 
(viewed as a planar density of electric dipoles)
by a homogeneously charged plate, and the moving charge by a time dependent
electric dipole $\vec d(t)$. In a sense this again corresponds to 
a transformation from the capacitor rest frame to that of 
the charge.
Hence the electric dipole effect is the dual  ``companion" of the 
potential AB effect.
The electric dipole experiences no forces, yet it acquires the
 phase
$ \phi_D  = {1\over \hbar} \int \vec d(t')\cdot \vec E dt' $. 
However viewing the dipole as formed by an  
extended (time dependent) charge distribution in an external electric
field, it will induce a corresponding  time dependent non-vanishing
internal stress.
The consistency of the potential AB effect, then requires
that the local internal effects should be complementary to the 
accumulated phase.
It would be interesting to understand the details of the complementarity
relation for this case and the related ``vector potential''  dipole effect.

We thank A. Casher, S. Nussinov and L. Vaidman, for helpful discussions.
We acknowledge the support from grant 471/98 of the Israel Science 
Foundation, established by the Israel Academy of Sciences and Humanities.
The work of Y. A. was supported by NSF grant PHY-9601280.

\noindent \dag {\em Email:\, reznik@post.tau.ac.il}

\end{document}